\documentclass{article}

\usepackage{amsfonts, amscd, amsmath, dsfont, clrscode}
\usepackage[dvips]{graphicx}

\begin{document}

\title{An algorithm for calculating steady state probabilities of $M|E_r|c|K$ queueing systems}
\author{Stefan Hochrainer \and Ronald Hochreiter \and Georg Pflug}
\date{July 2004}

\maketitle

\begin{abstract}
This paper presents a method for calculating steady state probabilities of $M|E_r|c|K$ queueing systems. The infinitesimal generator matrix is used to define all possible states in the system and their transition probabilities. While this matrix can be written down immediately for many other $M|PH|c|K$ queueing systems with phase-type service times (e.g. Coxian, Hypoexponential, \ldots), it requires a more careful analysis for systems with Erlangian service times. The constructed matrix may then be used to calculate steady state probabilities using an iterative algorithm. The resulting steady state probabilities can be used to calculate various performance measures, e.g. the average queue length. Additionally, computational issues of the implementation are discussed and an example from the field of telecommunication call-center queue length will be outlined to substantiate the applicability of these efforts. In the appendix, tables of the average queueing length given a specific number of service channels, 
traffic density, and system size are presented. 
\end{abstract}

\section{Introduction}

Multi-server queueing systems with Poisson input and phase-type distributed service times ($M|PH|c$ queueing systems) are an important extension of simple $M/M/c$ queueing systems\footnote{using the notation suggested by Kendall \cite{Kendall1}}. Different phase-type distributions are used for different fields of application.

The class of Erlang($r$) distributed service times ($M|E_r|c$) are important, because for constant mean service time the family of Erlangian distributions interpolates infinitely many distributions between the negative exponential $(r = 1)$ and constant service time $(r = \infty)$. Many solution approaches for different types of Erlangian distributed queueing systems have been proposed in the literature. Shapiro \cite{Shapiro1} studied the $M/E_2/c$ system and proved that the probability of $n$ jobs in the system can be expressed as a linear combination of powers. Poyntz and Jackson \cite{Poyntz1} analyzed the $E_k/E_r/2$ and the $E_k/E_r/3$ system by applying generating functions. Tables of such results have been published by Sakasegawa \cite{Sakasegawa1}. The $M/E_r/c$ system was analyzed by Mayhugh and McCormick \cite{MayMc1} and Heffer \cite{Heffer1}. They determined the generating function of the stationary state probabilities. However, their results are from a computational point of view only useful 
for small values of $r$ and $c$ only. Yu \cite{Yu1} solved the $E_k/E_r/c$ queueing system with heterogeneous servers by generalizing the approach of Mayhugh and McCormick \cite{MayMc1}. Hillier and Lo \cite{Hillier1} presented some numerical results based on the above procedure for the special case of homogeneous servers. Approximation formulas for the average waiting time and average system size in equilibrium were given by Page \cite{Page1} and Smith \cite{Smith1}. Hillier and Lo also presented tables and graphs of performance measures for different queueing systems. However, steady state probabilities for these queueing systems with limited waiting room are not given yet.

This paper presents an easy to implement algorithm to calculate the steady state probabilities of $M/E_r/c/K$ queueing systems. For this, we use a similar approach like Mayhugh and McCormick \cite{MayMc1}. Firstly, all possible states in the queueing system are determined and a generator matrix is built up. This generator matrix determines - via a well known theorem - the transition matrix. This transition matrix is used to find the steady state probabilities for being in a particular stage in the queueing system. These can be summed up to get the steady state probabilities for a specific number of customers in the queueing system. These probabilities can then be used to calculate different performance measures like the average queueing length and, with the help of Little's theorem \cite{Kleinrock1}, the average waiting time in the queue. In contrast to many other queueing systems with phase-type distributed service times, where the generator matrix can be written down immediately, the calculation of the 
matrix for $M/E_r/c/K$ systems needs a more detailed description, and is the main topic of this paper.

This paper is organized as follows. Section \ref{general} provides an overview of $M/E_r/c/K$ queueing systems and summarizes the algorithms used to calculate the generator matrix and derive steady state probabilities. Section \ref{example} gives an example how this method can be used in telecommunication call-centers to calculate performance measures and discusses computational issues of the algorithm presented above. Additionally, Appendix \ref{AppendixA} contains tables for the average queueing length with different traffic densities for the $M/E_r/c/K$ system.

\section{The $M|E_r|c|K$ queueing system}
\label{general}

\subsection{Assumptions}

The $M/E_r/c/K$ queueing system is defined by the following assumptions:

\begin{enumerate}
\item The arrival process of customers follows a Poisson distribution with intensity rate $\lambda$.
\item The population of customers is infinite.
\item The service system consists of $c$ service channels.
\item Only one customer can be served in one service channel.
\item The maximum queue size is $K$.
\item If an arriving customer finds all service channels busy and the maximum queue size is not reached, she joins the single waiting line which is served in the order of arrival.
\item An arriving customer is forced to balk if she arrives at a time when the queue size is at its limit $K$.
\item If an arriving customer finds more than one service channel vacant, she randomly selects a free service channel.
\item The customer at the head of the waiting line is forwarded to the first vacant channel without delay, i.e. the queue discipline is FCFS.
\item The service process of customers follows an Erlang probability distribution of degree $r$. The distribution of service time is equal in every channel.
\end{enumerate}

\subsection{Building the generator matrix}

The probability density of an Erlang($r$) distributed service time $\theta(t)$ is

\begin{equation}
\label{densservtime} 
f(t)=\mu(\mu t)^{r-1} \frac{e^{-\mu t}}{(r-1)!}
\end{equation}

Each service channel may be regarded as consisting of $r$ ordered stages such that the conditional probability of transition of a customer from any stage to the succeeding stage in the time interval $(t, t+\Delta t)$ is $\mu \Delta t + o(\Delta t)$. Equivalently, the distribution (\ref{densservtime}) can be seen as the distribution of the sum of $r$ independent negative exponentially distributed random variables each with the same parameter $\mu$. The reason for using Erlang distributed random variables in Queuing theory lies in the fact, that the phase method can be used to describe the process as a function of a Markov process. The cost is an enlarged state space for the model, and consequently an increased complexity of the numerical solution. The benefit, of course, is in enlarging the class of service time distributions for which the model is solvable. 

If the service time is Erlang($r$) distributed (\ref{densservtime}), the mean service time is $\frac{r}{\mu}$ and the mean traffic density per service channel in an $M/E_r/c$ system is thus

\begin{equation}
\rho=\frac{\lambda r}{\mu c}.
\end{equation}

$\frac{\lambda}{\mu}$ can be regarded as the mean traffic density per stage. It is well known in queueing theory \cite{Asmussen1} that a steady state solution exists if and only if $\rho < 1$. This means e.g. for queueing systems with limited waiting room and $\rho \geq 1$ that the average queueing length is near or at its maximum.

The transition matrix of a continuous-time Markov chain can be expressed \cite{Pflug1} as

\begin{equation}\label{transitionmatrix}
P(h) = \mbox{exp}(hQ)
\end{equation}

where $Q$ is the intensity matrix $Q = \lim_{h \downarrow 0} \frac{P(h) - I}{h}$. 

For an ergodic process, the limit of the transition matrix

\begin{equation}
\lim_{h \to \infty}P(h)=\lim_{h \to \infty} \exp(hQ)
\end{equation}

exists, is of rank 1 and its identical rows $\pi$ coincide with the stationary distribution of the Markov process. Instead of solving the linear equations $\pi Q = 0$, $\sum_i \pi_i = 1$, numerical solutions also can be obtained by using the generator matrix, formula \ref{transitionmatrix} and taking the power of $P$ until no changes in the rows occur anymore. 

We use this method to calculate the stationary probability distributions of the $M/E_r/c/K$ queueing system with the assumptions made above. First, we define the different states with the vector $s=(s_0,s_1,...,s_r)$ where $s_0$ is the queueing length and $s_i$ $(i=1,..,r)$ is the total number of customers being in phase $i$, this is a similar approach as in Mayhugh and McCormick \cite{MayMc1}.

The total number of possible states can be calculated through the equation:
\begin{equation}
\label{nopossiblestates}
N = \sum_{i=1}^{c}\dbinom{i+r-1}{i} + \dbinom{c+r-1}{c} +1
\end{equation}
or in closed form, calculated with the Zeilberger algorithm
$$N = \frac{c+2r}{r} \dbinom{c+r-1}{c}$$
where the first part in (\ref{nopossiblestates}) is the number of states where no customers are waiting $(s_0=0)$ and the second part is the number of states where customers are waiting to get served $(s_0>0)$. The third part refers to the empty state $s_i=0$ for $i=0,...,r$. A possible algorithm to calculate these states as matrix $M_s$ is given in Algorithm 1. 

\begin{table}
\begin{tabular}{p{11cm}} \hline
Algorithm 1. Initial Matrix Setup \\ \hline
\end{tabular}
\begin{codebox}
\Procname{$\proc{MatrixSetup}(k, c)$}
\li $t \gets 1$
\li $z \gets \proc{zeros}(k)$
\li \While $z_k \leq c$
\li	\>  \If $\sum z \leq c$ \kw{then}
\li \> \> $M_s(t, 2:k+1) \gets z$
\li \> \>             $t \gets t + 1$
\li \> \kw{end if} 
\li \>	  $z_1 \gets z_1 + 1$
\li \>  \For $i \gets 2$ \To $k$
\li \> \> \If $z_{i-1} > c$ \kw{then} 
\li \> \> \> $z_{i-1} \gets 0$
\li \> \> \> $z_i \gets z_i + 1;$
\li \> \> \kw{end if} 
\li \> \kw{next} 
\li \kw{end while} 
\end{codebox}
\hrule 
\end{table}

It should be clear that other algorithms, like the lexicographically ordering in Mayhugh and McCormick \cite{MayMc1}, could also be used. However, the states created here are the same, but in different orders. We note, that we only have to find all possible states for a specific queuing system and it is not necessary to order the states in a specific way to build up our generating matrix.

After that, we have to create all possible states where customers are waiting. The number of states in which $j, (j=1,...,K)$ customers are waiting is independent from $j$ and therefore a constant equal to $\dbinom{c+r-1}{c}$. Furthermore, the set of possible states for different numbers of waiting customers are equal except in $s_0$. E.g. the set of states with $s_0=1$ are identical to the set of states with $s_0=j, (j=2,...,K)$ except for the number of customers in the queue. We can use the created states from the algorithm above, to find all possible states for the queuing system with waiting customers. Therefore, we are looking only at the created states where the $\sum_{i=1}^{r}s_i=c$, because in this situation arriving customers could not be served and have to wait in the queue. To create all possible states with $s_0=1$ we only have to take the states created by the above algorithm with $s_0=0$ and $\sum_{i=1}^{r}s_i=c$ and set $s_0=1$ instead of $s_0=0$, the same procedure can be used for $s_0=2$ 
until $s_0=K$. 

Because of (\ref{nopossiblestates}) the generator matrix must have size $N \times N$. Below the possible transitions from one state to another are summarized:

\begin{enumerate}
\item If the queue is empty ($s_0=0$) and at least one service channel is free ($\sum_{i=1}^{r}s_i <c$) an arriving customer starts  phase 1 (from ($s_0,s_1,...,s_r$) to ($s_0,s_1+1,...,s_r$) with intensity $\lambda$).
\item If the queue is not at its maximum ($0\leq s_0 <K$) and all service channels are busy ($\sum_{i=1}^{r}s_i =c$) the arriving customer joins the queue (from ($s_0,s_1,...,s_r$) to ($s_0+1,s_1,...,s_r$) with intensity $\lambda$)..
\item A customer being in phase $s_i$ $i=1,...,r-1$ transits to phase $s_{i+1}$ $i=2,...,r$ with intensity $s_i\mu$.
\item If a customer is in phase $s_r$ she leaves the system with intensity $\mu$. If there is a customer waiting in the queue she starts in phase $s_1$, otherwise the server is idle until the next customer arrives.
\end{enumerate}

The rows of the matrix Q can be seen as the starting point of each state and the columns are filled in the way described above. After that, the $q_{ii}$ ($i=1,...,N$) can be calculated by the formula:

\begin{equation}
q_{ii}=-\sum_{i \neq j}q_{ij}
\end{equation}

\paragraph{Matrix Setup Example}

As an example we consider the $M/E_2/2/1$ queue. First we give a table of all possible states for this system, it was generated with the algorithm given above:

\begin{center}
\begin{tabular}{|l|lllllllll|} \hline
$s_0$ & 0 & 0 & 0 & 0 & 0 & 0 & 1 & 1 & 1 \\ \hline
$s_1$ & 0 & 1 & 2 & 0 & 1 & 0 & 2 & 1 & 0 \\ \hline 
$s_2$ & 0 & 0 & 0 & 1 & 1 & 2 & 0 & 1 & 2 \\ \hline
\end{tabular}
\end{center}

The generator matrix build with the algorithm above is

\begin{equation*}
\begin{pmatrix}
-\lambda &     \lambda  &   0   &  0 &    0   &  0    & 0 &    0 &    0\\
     0&    -(\lambda+\mu) &    \lambda  &   \mu   &  0  &   0   &  0  &   0 &    0\\
     0&     0&    -(\lambda+2\mu) &    0   &  2\mu  &   0  &   \lambda &    0 &    0\\
     \mu&     0 &    0  &  -(\lambda+\mu)&     \mu &    0  &   0   &  0  &   0\\
     0&     \mu &     0 &    0 &   -(\lambda +2\mu)  &   \mu  &   0 &    \lambda &    0\\
     0&     0 &    0  &   2\mu &    0 &   -(\lambda+2\mu)  &   0 &    0 &    \lambda\\
     0&     0 &    0 &    0 &    0  &   0   & -2\mu  &   2\mu  &   0\\
     0 &    0&     \mu &    0  &   0 &    0  &   0   & -2\mu &    \mu\\
     0&     0&     0 &    0  &   2\mu  &   0  &   0   &  0 &   -2\mu\\
\end{pmatrix}
\end{equation*}

The first row represents the state (0,0,0), as we can see the only transition to another state is given by (0,1,0) with intensity $\lambda$. The sixth row represents the state (0,0,2), this state can go to (0,0,1) with intensity $2 \mu$ and to (1,0,2) with intensity $\lambda$.

\subsection{Calculating steady state probabilities}

The algorithm for calculating steady state probabilities of the $M/E_r/c/K$ queueing system needs five steps and an optional sixth step:

\begin{enumerate}
\item{Calculate all possible states}
\item{Generate the $Q$-matrix with the steps explained above}
\item{Use formula (5) to get the transition matrix}
\item{Take the power of the transition matrix iteratively until the columns don't change anymore (see Algorithm 2)}
\item{Take the sum of the different steady state probabilities for the stages to get the steady state probabilities for $n$ customers in the system}
\item{(Optional) Calculate performance measures}
\end{enumerate}

Step 4 is needed to calculate the steady state probabilities for the different stages. Let $P$ be the transition matrix, a possible algorithm could be a simple iteration like shown in algorithm 2 where $\delta$ is some critical level defined by the user.
\begin{table}
\begin{tabular}{p{11cm}} \hline
Algorithm 2. Iterative Transition Matrix Calculation \\ \hline
\end{tabular}
\begin{codebox}
\Procname{$\proc{IterativeTransitionMatrix}(P)$}
\li $P_{old} \gets P$
\li $P_{new} \gets (P)^2$
\li \While ($P_{old} - P_{new}) > \delta$ \kw{do}
\li \> $P_{old} \gets P_{new}$
\li \> $P_{new} \gets (P_{new})^2$
\li \kw{end while} 
\end{codebox}
\hrule
\end{table}
Step 5 in the algorithm above can be calculated in various ways. The chosen method depends on the algorithm which is used to create the possible states. If an algorithm is used which creates states like in Mayhugh and McCormick \cite{MayMc1} formulas to calculate the probability of $n$ customers in the system $P_n$ can be devised easily. However, to calculate the possible states with the algorithm presented here, a different formula has to be used. Denote the state probabilities with $p_j$. We first notice that $P_0 = p_0$, and if $n > c$ then $P_n$ can be calculated with the formula given in Mayhugh and McCormick \cite{MayMc1}, pp. 710. For $P_n$ with $0 < n \leq r$ we have to sum up the states $p_j$ for which $\sum_{i=1}^r s_i = n$.

Step 6 is optional, if some performance measures are requested, they can be easily calculated with the steady state probabilities $P_n$, see for example Gross and Harris \cite{Gross1}. E.g. the average system size can be calculated, using the well known formula:

\begin{equation*}
L=\sum_{n=0}^{c+K} nP_n
\end{equation*}

\subsection{Empirical computational issues}

\begin{figure}
\begin{center}
\scalebox{0.6}{\includegraphics{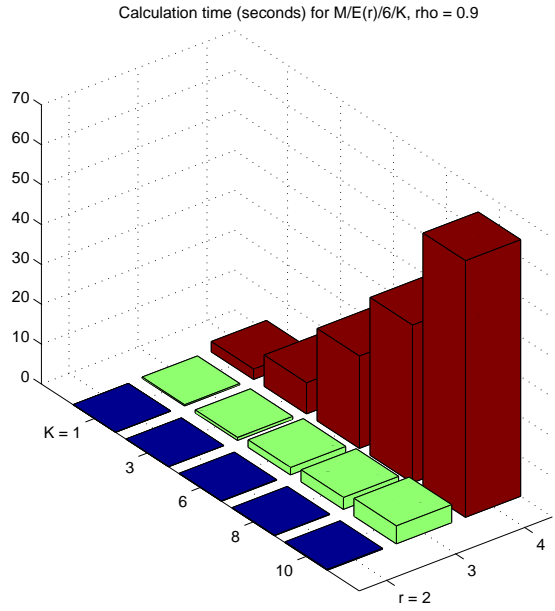}}
\end{center}
\caption{Calculation time of the average queueing length}
\label{fig:queuecalc1}
\end{figure}

The current algorithm is easy to implement, but a large waiting queue $K$ with more than $4$ phases leads to high computational times. The following run-time experiments have been conducted on a Pentium IV computer with 2.6 Ghz and 1 GB RAM using Microsoft Windows XP Professional and MatLab 6.5. Figure \ref{fig:queuecalc1} shows the calculation time in seconds for average queueing length of a $M/E_r/c/K$ system with $\rho = 0.9$, $c = 6$ for $r = 2, \ldots, 4$ and $K \in \{ 1, 3, 6, 8, 10 \}$.

\section{Telecommunication Call-Centers}
\label{example}

\begin{figure}
\begin{center}
\scalebox{0.6}{\includegraphics{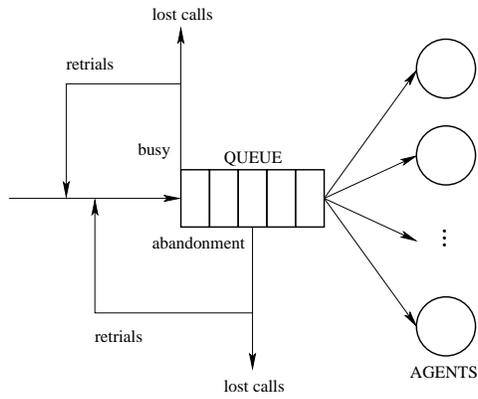}}
\end{center}
\caption{General Call-Center Queue}
\label{figccqueue1}
\end{figure}

\begin{figure}
\begin{center}
\scalebox{0.6}{\includegraphics{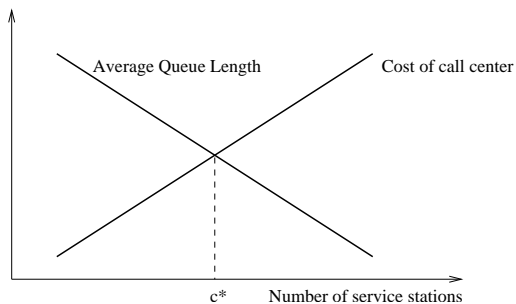}}
\end{center}
\caption{Schematical representation for a management decision about the number of service stations $c^*$}
\label{fig:cccost1}
\end{figure}

A call center is a service network in which agents provide telephone-based services. The typical call center setup is shown in figure \ref{figccqueue1}. When all agents are busy customers seeking these services are delayed in queues, hence it is convenient to model call centers as a queueing system. Process-wise these queues can be compared to inventories in manufacturing (just-in-time, time-based-competition, \ldots). But human queues include personal preferences, complaints, abandonments and the like. Thus, customers are likely to base judgments about the service-providing company on their queueing-experience. Therefore the goal of a company providing such a service is to minimize the average waiting time of their costumers. This can always be accomplished by extending the number of agents, which in turn raises the cost of the call center significantly. The decision problem is schematically shown in figure \ref{fig:cccost1}. To take a good decision $c^*$ it is important to calculate the average queue 
length as correct as possible.

A survey of queueing system theory for call centers was published by Koole and Mandelbaum \cite{Koole02} and more detailed by Koole et al. \cite{Koole03}. Recently Ishay \cite{Ishay03} applied phase type distributions to fit call center data and found out that phase type distributions of order $k = 2, 3, 4, 5, 6$ can be used to fit the service durations of call-center data for different priorities and service-types. The general structure of order $k = 3$ already provides a reasonable fit to the overall service time. She used the program EMpht (see S. Asmussen et al. \cite{Asmussen96}) to fit phase-type distributions to available data from a call center of a large Israelian bank. 

\section{Conclusion}
\label{conclusion}

In this paper we presented an easy-to-implement algorithm for calculating steady state probabilities of $M/E_r/c/K$ queueing systems. In contrast to other types of queueing systems with phase-type distributed service times, the main problem with the type of queueing system considered throughout this paper is how to generate the generator matrix. Hence, an algorithm to generate this matrix was presented. The resulting methodology can be used for several practical applications, including telecommunication call-centers. Phase-type distributions and associated average system size calculations are well suited to re-design and re-dimension existing call centers.

Future research includes an extension of this methodology to $PH|PH|c$ (especially $E_s|E_r|c|K$) queueing systems.

\begin{appendix}
\section{Tables}
\label{AppendixA}


\begin{table}
\begin{tabular}{|c|c|ccccccccc|} \hline
\multicolumn{2}{|c|}{} & \multicolumn{9}{c|}{$\rho$} \\ \cline{3-11}
\multicolumn{2}{|c|}{} & 0.1 & 0.3 & 0.5 & 0.7 & 0.8 & 0.9 & 0.95 & 0.98 & 0.99 \\ \hline
  & 1  & 0.400 & 1.198 & 1.958 & 2.606 & 2.876 & 3.110 & 3.214 & 3.274 & 3.293 \\
  & 3  & 0.400 & 1.211 & 2.090 & 3.051 & 3.528 & 3.977 & 4.184 & 4.303 & 4.342 \\
K & 5  & 0.400 & 1.212 & 2.132 & 3.384 & 4.190 & 5.071 & 5.511 & 5.768 & 5.852 \\
  & 7  & 0.400 & 1.212 & 2.136 & 3.480 & 4.472 & 5.683 & 6.325 & 6.707 & 6.832 \\
  & 10 & 0.400 & 1.212 & 2.136 & 3.527 & 4.664 & 6.212 & 7.090 & 7.622 & 7.797 \\ \hline
\end{tabular}
\caption{Average system size $M|E_2|4|K$}
\end{table}

\begin{table}
\begin{tabular}{|c|c|ccccccccc|} \hline
\multicolumn{2}{|c|}{} & \multicolumn{9}{c|}{$\rho$} \\ \cline{3-11}
\multicolumn{2}{|c|}{} & 0.1 & 0.3 & 0.5 & 0.7 & 0.8 & 0.9 & 0.95 & 0.98 & 0.99 \\ \hline
  & 1  & 0.600 & 1.798 & 2.952 & 3.941 & 4.346 & 4.689 & 4.840 & 4.924 & 4.951 \\
  & 3  & 0.600 & 1.803 & 3.046 & 4.348 & 4.979 & 5.559 & 5.824 & 5.974 & 6.022 \\
K & 5  & 0.600 & 1.803 & 3.076 & 4.646 & 5.612 & 6.648 & 7.159 & 7.455 & 7.552 \\
  & 7  & 0.600 & 1.804 & 3.078 & 4.731 & 5.879 & 7.253 & 7.975 & 8.401 & 8.540 \\
  & 10 & 0.600 & 1.804 & 3.079 & 4.773 & 6.060 & 7.775 & 8.739 & 9.320 & 9.511 \\ \hline
\end{tabular}
\caption{Average system size $M|E_2|6|K$}
\end{table}

\begin{table}
\begin{tabular}{|c|c|ccccccccc|} \hline
\multicolumn{2}{|c|}{} & \multicolumn{9}{c|}{$\rho$} \\ \cline{3-11}
\multicolumn{2}{|c|}{} & 0.1 & 0.3 & 0.5 & 0.7 & 0.8 & 0.9 & 0.95 & 0.98 & 0.99 \\ \hline
  & 1  & 0.800 & 2.399 & 3.957 & 5.309 & 5.857 & 6.315 & 6.512 & 6.621 & 6.656 \\
  & 3  & 0.800 & 2.401 & 4.025 & 5.681 & 6.470 & 7.184 & 7.505 & 7.685 & 7.743 \\
K & 5  & 0.800 & 2.401 & 4.045 & 5.950 & 7.078 & 8.268 & 8.848 & 9.181 & 9.290 \\
  & 7  & 0.800 & 2.401 & 4.047 & 6.026 & 7.332 & 8.867 & 9.665 & 10.13 & 10.28 \\
  & 10 & 0.800 & 2.401 & 4.048 & 6.062 & 7.503 & 9.382 & 10.42 & 11.05 & 11.26 \\ \hline
\end{tabular}
\caption{Average system size $M|E_2|8|K$}
\end{table}

\begin{table}
\begin{tabular}{|c|c|ccccccccc|} \hline
\multicolumn{2}{|c|}{} & \multicolumn{9}{c|}{$\rho$} \\ \cline{3-11}
\multicolumn{2}{|c|}{} & 0.1 & 0.3 & 0.5 & 0.7 & 0.8 & 0.9 & 0.95 & 0.98 & 0.99 \\ \hline
  & 1  & 1.000 & 2.999 & 4.965 & 6.696 & 7.395 & 7.971 & 8.216 & 8.350 & 8.393 \\
  & 3  & 1.000 & 3.000 & 5.013 & 7.035 & 7.987 & 8.837 & 9.215 & 9.425 & 9.492 \\
K & 5  & 1.000 & 3.000 & 5.028 & 7.279 & 8.571 & 9.915 & 10.56 & 10.93 & 11.05 \\
  & 7  & 1.000 & 3.000 & 5.029 & 7.347 & 8.815 & 10.50 & 11.38 & 11.88 & 12.05 \\
  & 10 & 1.000 & 3.000 & 5.029 & 7.380 & 8.977 & 11.01 & 12.14 & 12.81 & 13.03 \\ \hline
\end{tabular}
\caption{Average system size $M|E_2|10|K$}
\end{table}

\begin{table}
\begin{tabular}{|c|c|ccccccccc|} \hline
\multicolumn{2}{|c|}{} & \multicolumn{9}{c|}{$\rho$} \\ \cline{3-11}
\multicolumn{2}{|c|}{} & 0.1 & 0.3 & 0.5 & 0.7 & 0.8 & 0.9 & 0.95 & 0.98 & 0.99 \\ \hline
  & 1  & 1.500 & 4.499 & 7.482 & 10.20 & 11.30 & 12.19 & 12.56 & 12.76 & 12.82 \\
  & 3  & 1.500 & 4.500 & 7.502 & 10.47 & 11.84 & 13.05 & 13.56 & 13.85 & 13.94 \\
K & 5  & 1.500 & 4.500 & 7.508 & 10.66 & 12.38 & 14.11 & 14.92 & 15.38 & 15.53 \\
  & 7  & 1.500 & 4.500 & 7.509 & 10.71 & 12.59 & 14.69 & 15.74 & 16.35 & 16.54 \\
  & 10 & 1.500 & 4.500 & 7.509 & 10.74 & 12.74 & 15.19 & 16.50 & 17.28 & 17.53 \\ \hline
\end{tabular}
\caption{Average system size $M|E_2|15|K$}
\end{table}


\begin{table}
\begin{tabular}{|c|c|ccccccccc|} \hline
\multicolumn{2}{|c|}{} & \multicolumn{9}{c|}{$\rho$} \\ \cline{3-11}
\multicolumn{2}{|c|}{} & 0.1 & 0.3 & 0.5 & 0.7 & 0.8 & 0.9 & 0.95 & 0.98 & 0.99 \\ \hline
  & 1  & 0.400 & 1.198 & 1.959 & 2.610 & 2.881 & 3.117 & 3.223 & 3.282 & 3.301 \\
  & 3  & 0.400 & 1.211 & 2.086 & 3.049 & 3.533 & 3.990 & 4.202 & 4.323 & 4.363 \\
K & 5  & 0.400 & 1.211 & 2.121 & 3.353 & 4.163 & 5.068 & 5.527 & 5.795 & 5.883 \\
  & 7  & 0.400 & 1.211 & 2.124 & 3.431 & 4.414 & 5.655 & 6.329 & 6.732 & 6.864 \\
  & 10 & 0.400 & 1.211 & 2.124 & 3.466 & 4.576 & 6.151 & 7.074 & 7.640 & 7.827 \\ \hline
\end{tabular}
\caption{Average system size $M|E_3|4|K$}
\end{table}

\begin{table}
\begin{tabular}{|c|c|ccccccccc|} \hline
\multicolumn{2}{|c|}{} & \multicolumn{9}{c|}{$\rho$} \\ \cline{3-11}
\multicolumn{2}{|c|}{} & 0.1 & 0.3 & 0.5 & 0.7 & 0.8 & 0.9 & 0.95 & 0.98 & 0.99 \\ \hline
  & 1  & 0.600 & 1.798 & 2.953 & 3.945 & 4.352 & 4.697 & 4.848 & 4.933 & 4.960 \\
  & 3  & 0.600 & 1.803 & 3.045 & 4.350 & 4.987 & 5.576 & 5.846 & 5.998 & 6.047 \\
K & 5  & 0.600 & 1.803 & 3.070 & 4.626 & 5.596 & 6.655 & 7.184 & 7.491 & 7.591 \\
  & 7  & 0.600 & 1.803 & 3.072 & 4.696 & 5.836 & 7.239 & 7.990 & 8.436 & 8.582 \\
  & 10 & 0.600 & 1.803 & 3.073 & 4.727 & 5.989 & 7.730 & 8.737 & 9.350 & 9.552 \\ \hline
\end{tabular}
\caption{Average system size $M|E_3|6|K$}
\end{table}

\begin{table}
\begin{tabular}{|c|c|ccccccccc|} \hline
\multicolumn{2}{|c|}{} & \multicolumn{9}{c|}{$\rho$} \\ \cline{3-11}
\multicolumn{2}{|c|}{} & 0.1 & 0.3 & 0.5 & 0.7 & 0.8 & 0.9 & 0.95 & 0.98 & 0.99 \\ \hline
  & 1  & 0.800 & 2.399 & 3.958 & 5.313 & 5.863 & 6.322 & 6.521 & 6.630 & 6.665 \\
  & 3  & 0.800 & 2.401 & 4.024 & 5.684 & 6.480 & 7.204 & 7.529 & 7.712 & 7.771 \\
K & 5  & 0.800 & 2.401 & 4.042 & 5.936 & 7.068 & 8.281 & 8.878 & 9.222 & 9.334 \\
  & 7  & 0.800 & 2.401 & 4.044 & 6.000 & 7.299 & 8.862 & 9.688 & 10.17 & 10.33 \\
  & 10 & 0.800 & 2.401 & 4.044 & 6.028 & 7.446 & 9.349 & 10.43 & 11.09 & 11.31 \\ \hline
\end{tabular}
\caption{Average system size $M|E_3|8|K$}
\end{table}

\begin{table}
\begin{tabular}{|c|c|ccccccccc|} \hline
\multicolumn{2}{|c|}{} & \multicolumn{9}{c|}{$\rho$} \\ \cline{3-11}
\multicolumn{2}{|c|}{} & 0.1 & 0.3 & 0.5 & 0.7 & 0.8 & 0.9 & 0.95 & 0.98 & 0.99 \\ \hline
  & 1  & 1.000 & 2.999 & 4.965 & 6.700 & 7.400 & 7.978 & 8.224 & 8.359 & 8.402 \\
  & 3  & 1.000 & 3.000 & 5.013 & 7.040 & 7.998 & 8.858 & 9.240 & 9.453 & 9.521 \\
K & 5  & 1.000 & 3.000 & 5.026 & 7.269 & 8.567 & 9.933 & 10.59 & 10.97 & 11.10 \\
  & 7  & 1.000 & 3.000 & 5.027 & 7.327 & 8.789 & 10.51 & 11.41 & 11.93 & 12.11 \\
  & 10 & 1.000 & 3.000 & 5.027 & 7.353 & 8.929 & 10.99 & 12.16 & 12.86 & 13.09 \\ \hline
\end{tabular}
\caption{Average system size $M|E_3|10|K$}
\end{table}


\begin{table}
\begin{tabular}{|c|c|ccccccccc|} \hline
\multicolumn{2}{|c|}{} & \multicolumn{9}{c|}{$\rho$} \\ \cline{3-11}
\multicolumn{2}{|c|}{} & 0.1 & 0.3 & 0.5 & 0.7 & 0.8 & 0.9 & 0.95 & 0.98 & 0.99 \\ \hline
  & 1  & 0.400 & 1.198 & 1.959 & 2.612 & 2.885 & 3.121 & 3.227 & 3.286 & 3.306 \\
  & 3  & 0.400 & 1.210 & 2.084 & 3.047 & 3.534 & 3.996 & 4.211 & 4.334 & 4.373 \\
K & 5  & 0.400 & 1.211 & 2.116 & 3.336 & 4.147 & 5.065 & 5.534 & 5.809 & 5.899 \\
  & 7  & 0.400 & 1.211 & 2.118 & 3.405 & 4.381 & 5.637 & 6.328 & 6.743 & 6.880 \\
  & 10 & 0.400 & 1.211 & 2.118 & 3.434 & 4.525 & 6.114 & 7.062 & 7.647 & 7.841 \\ \hline
\end{tabular}
\caption{Average system size $M|E_4|4|K$}
\end{table}

\begin{table}
\begin{tabular}{|c|c|ccccccccc|} \hline
\multicolumn{2}{|c|}{} & \multicolumn{9}{c|}{$\rho$} \\ \cline{3-11}
\multicolumn{2}{|c|}{} & 0.1 & 0.3 & 0.5 & 0.7 & 0.8 & 0.9 & 0.95 & 0.98 & 0.99 \\ \hline
  & 1  & 0.600 & 1.798 & 2.953 & 3.947 & 4.355 & 4.701 & 4.853 & 4.937 & 4.965 \\
  & 3  & 0.600 & 1.803 & 3.044 & 4.350 & 4.991 & 5.585 & 5.857 & 6.010 & 6.060 \\
K & 5  & 0.600 & 1.803 & 3.068 & 4.614 & 5.585 & 6.657 & 7.195 & 7.509 & 7.611 \\
  & 7  & 0.600 & 1.803 & 3.069 & 4.677 & 5.811 & 7.227 & 7.996 & 8.454 & 8.604 \\
  & 10 & 0.600 & 1.803 & 3.069 & 4.703 & 5.949 & 7.701 & 8.732 & 9.364 & 9.573 \\ \hline
\end{tabular}
\caption{Average system size $M|E_4|6|K$}
\end{table}

\begin{table}
\begin{tabular}{|c|c|ccccccccc|} \hline
\multicolumn{2}{|c|}{} & \multicolumn{9}{c|}{$\rho$} \\ \cline{3-11}
\multicolumn{2}{|c|}{} & 0.1 & 0.3 & 0.5 & 0.7 & 0.8 & 0.9 & 0.95 & 0.98 & 0.99 \\ \hline
  & 1  & 0.800 & 2.399 & 3.958 & 5.315 & 5.866 & 6.326 & 6.525 & 6.635 & 6.670 \\
  & 3  & 0.800 & 2.401 & 4.024 & 5.686 & 6.485 & 7.213 & 7.542 & 7.726 & 7.785 \\
K & 5  & 0.800 & 2.401 & 4.041 & 5.928 & 7.061 & 8.286 & 8.893 & 9.243 & 9.357 \\
  & 7  & 0.800 & 2.401 & 4.042 & 5.986 & 7.279 & 8.856 & 9.698 & 10.19 & 10.36 \\
  & 10 & 0.800 & 2.401 & 4.042 & 6.009 & 7.413 & 9.327 & 10.43 & 11.11 & 11.33 \\ \hline
\end{tabular}
\caption{Average system size $M|E_4|8|K$}
\end{table}

\end{appendix}

\bibliography{hhp-merck}

\begin{thebibliography}{10}

\bibitem{Asmussen1}
S.~Asmussen.
\newblock {\em Applied probability and queues}, volume~51 of {\em Applications
  of Mathematics}.
\newblock Springer, 2nd edition, 2003.

\bibitem{Asmussen96}
S.~Asmussen, O.~Nerman, and M.~Olsson.
\newblock Fitting phase-type distribution via the {EM} algorithm.
\newblock {\em Scandinavian Journal of Statistics}, 23:419--441, 1996.

\bibitem{Koole03}
N.~Gans, G.~Koole, and A.~Mandelbaum.
\newblock Telephone call centers: Tutorial, review, and research prospects.
\newblock {\em Manufacturing \& Service Operations Management}, 5:79--141,
  2003.

\bibitem{Gross1}
D.~Gross and C.~M. Harris.
\newblock {\em Applied probability and queues}.
\newblock Wiley, New York, 1974.

\bibitem{Heffer1}
J.~C. Heffer.
\newblock Steady state solution of the ${M}/{E}_k/c(\infty, \mbox{FIFO})$
  queueing system.
\newblock {\em INFOR J. Canadian O.R.S.}, 17:16--30, 1969.

\bibitem{Hillier1}
F.S. Hillier and F.D. Lo.
\newblock Tables for multi-server queueing systems involving erlang
  distributions.
\newblock Technical Report~14, Operations Research Department, Stanford
  University, 1971.

\bibitem{Ishay03}
E.~Ishay.
\newblock Fitting phase-type distributions to data from a telephone call
  center.
\newblock Master's thesis, Technion - Israel Institute Of Technology, 2003.

\bibitem{Kendall1}
D.~Kendall.
\newblock Some problems in the theory of queue.
\newblock {\em Journal of the Royal Statistical Society}, 13:151--153, 1951.

\bibitem{Kleinrock1}
L.~Kleinrock.
\newblock {\em Queueing systems, {V}olume 1: {T}heory}.
\newblock Wiley, New York, 1975.

\bibitem{Koole02}
G.~Koole and A.~Mandelbaum.
\newblock Queueing models of call centers: An introduction.
\newblock {\em Annals of Operations Research}, 113:41--59, 2002.

\bibitem{MayMc1}
J.~O. Mayhugh and R.~E. McCormick.
\newblock Steady-state solution of the ${M}/{E}_k/r$.
\newblock {\em Management Science}, 14:692--712, 1968.

\bibitem{Page1}
E.~Page.
\newblock Tables of waiting time for ${M}/{M}/n,{M}/{D}/n$ and ${D}/{M}/n$ and
  their use to give approxiate waiting times in more general queues.
\newblock {\em J. Op. Res. Soc.}, 33:453--473, 1982.

\bibitem{Pflug1}
G.~Ch. Pflug.
\newblock {\em Stochastische Modelle in der Informatik}.
\newblock Stuttgart, 1986.

\bibitem{Poyntz1}
C.D. Poyntz and R.R.P. Jackson.
\newblock The steady-state solution for the queuing process ${E}_k/{E}_m/r$.
\newblock {\em Operations Research Quarterly}, 24:615--625, 1973.

\bibitem{Sakasegawa1}
H.~Sakasegawa.
\newblock Numerical tables of the queueing systems 1:${E}_k/{E}_2/s$.
\newblock Institute of Statistical Mathematics, Computer Science Monograph,
  1978.

\bibitem{Shapiro1}
S.~Shapiro.
\newblock The {M}-server queue with poisson input and gamma distributed service
  of order two.
\newblock {\em Operations Research}, 14:685--694, 1966.

\bibitem{Smith1}
V.~L. Smith.
\newblock Approximating the distribution of customers in ${M}/{E}_n/s$ queues.
\newblock {\em J. Op. Res. Soc.}, 36:327--332, 1985.

\bibitem{Yu1}
O.S. Yu.
\newblock On the steady-state solution of an ${E}_m/{E}_k/r$ queue with
  heterogeneous servers.
\newblock Technical Report~38, Operations Research Department, Stanford
  University, 1971.

\end{thebibliography}
\bibliographystyle{plain}

\end{document}